\documentclass[floats,aps,prb,showpacs,superscriptaddress]{revtex4}
\usepackage{graphicx}
\usepackage{amsmath}
\usepackage{natbib}
\usepackage{subfig}

\begin{document}
\title{Spin injection from topological insulator tunnel-coupled to
  metallic leads}

\author{P. P. Aseev}
% \thanks{Email address: art@cplire.ru}}
\affiliation{Kotel'nikov Institute of Radio-engineering and
  Electronics of Russian Academy of Sciences, Moscow 125009, Russia}
\affiliation{Moscow Institute of Physics and Technology, Dolgoprudny
  141700, Moscow region, Russia}

\author{S. N. Artemenko} \email[E-mail: ]{art@cplire.ru}
\affiliation{Kotel'nikov Institute of Radio-engineering and
  Electronics of Russian Academy of Sciences, Moscow 125009, Russia}
\affiliation{Moscow Institute of Physics and Technology, Dolgoprudny
  141700, Moscow region, Russia} \date{\today}

\begin{abstract}
  We study theoretically helical edge states of 2D and 3D topological
  insulators (TI) tunnel-coupled to metal leads and show that their
  transport properties are strongly affected by contacts as the latter
  play a role of a heat bath and induce damping and relaxation of
  electrons in the helical states of TI. A simple structure that
  produces a pure spin current in the external circuit is
  proposed. The current and spin current delivered to the external
  circuit depend on relation between characteristic lengths: damping
  length due to tunneling, contact length and, in case of 3D TI, mean
  free path and spin relaxation length caused by momentum
  scattering. If the damping length due to tunneling is the smallest
  one, then the electric and spin currents are proportional to the
  conductance quantum in 2D TI, and to the conductance quantum
  multiplied by the ratio of the contact width to the Fermi wavelength in
  3D TI.
\end{abstract}

\maketitle

% Introduction
Spin properties of edge and surface states of topological insulators
(TI) are of great interest both for fundamental physics and for
potential applications in spintronics~\cite{Pesin2012spintronics}. The
spin of electrons is strongly coupled to their momentum giving an idea
of generating spin polarized currents in TI~\cite{ModakPRB2012,
  BruneNatPhys2012, SukhanovSablikovJPhys2012}. However, it would be
interesting and of practical importance to generate not only spin
polarized currents but pure spin currents as well. The general idea
for generating pure spin current was suggested in
Ref.~\onlinecite{PareekPRL2004}: a Y-shaped two-dimensional conductor
forming a three-terminal junction with intrinsic spin-orbit
interaction was proposed, where one of the terminals is a voltage
probe which draws no electric current, but the polarizations of
incoming and outcoming electrons are opposite to each other, causing a
pure spin current. However, the particular realization of this system does
not relate to TI.
An example of a multiterminal system involving the edge state of TI,
in which a pure spin current in the external circuit may occur is
given in Ref.~\onlinecite{DolciniPRB2011}. However, the decoherence
and damping induced by contacts were out of consideration, while we
find that damping and relaxation induced by coupling to a metallic
contact are very important. The systems for generating a pure spin
current suggested in Refs.~\onlinecite{PareekPRL2004,DolciniPRB2011}
were mesoscopic and
ballistic. 
It is interesting to study a possibility to produce a pure spin current
also in a 3D TI where the spin current can be larger as it is
proportional to geometrical dimensions of the sample. In the helical
surface state of 3D TI the physics is more complicated because a finite angle impurity
scattering is not prohibited by momentum-spin locking and strongly affects transport properties.

In this paper we study an edge state in a 2D TI and a surface state in
3D TI coupled to metallic leads by tunnel contacts, take into account
decoherence due to exchange of electrons with the lead and due to
impurity scattering in 3D TI, and calculate charge and spin currents
in the external circuit. A distinctive feature of our approach is that
we take into account the decoherence induced by the contacts and show that
it determines the electric and spin currents in the TI with
contacts. We find that the currents strongly depend on relations
between the characteristic lengths: the damping length due to tunneling,
the length of the contact and the mean free path.

Below we set $e$, $\hbar$ and $k_B$ to unity, restoring dimensional
units in final expressions when necessary.

% Problem formulation

\begin{figure}
  \subfloat[]{\label{fig:system-edge}\includegraphics[width=5.3cm]{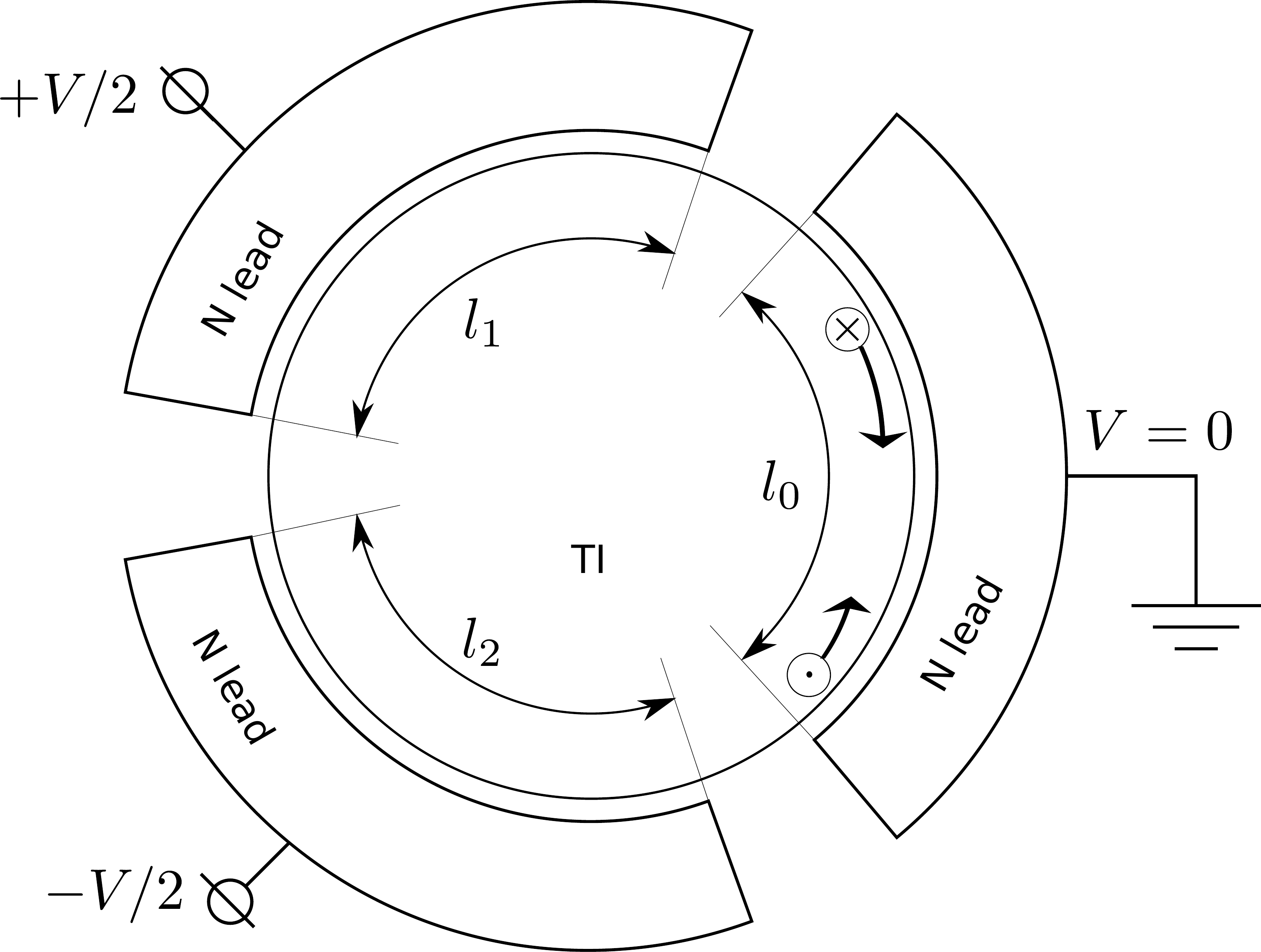}}\quad
  \subfloat[]{\label{fig:system-surf}\includegraphics[width=2.8cm]{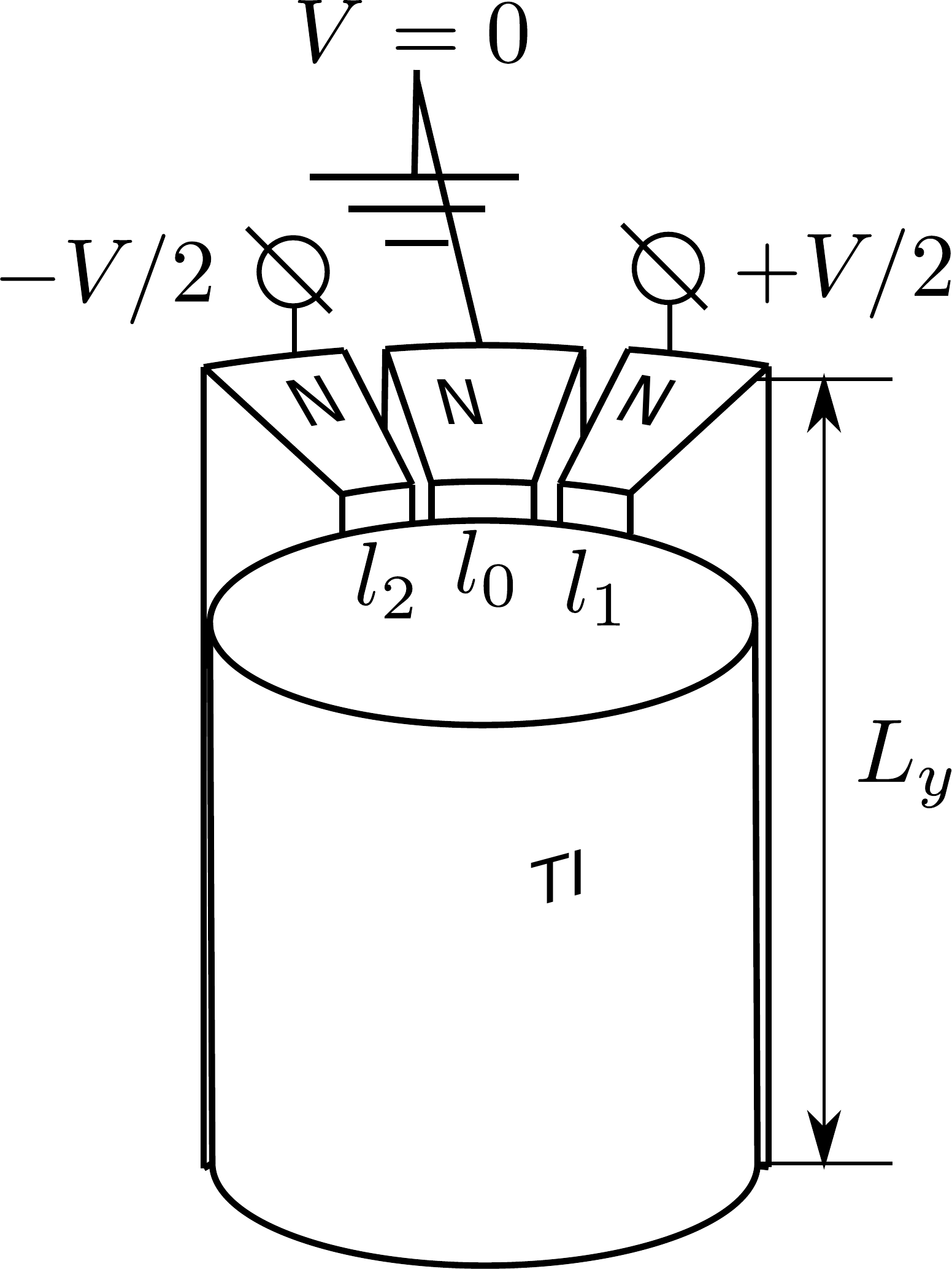}}
  \caption{(a) Helical edge state of 2D TI coupled to the leads (b)
    Helical surface state of 3D TI coupled to the leads}
  \label{fig:system}
\end{figure}

We consider a TI with a conducting helical state coupled by tunnel
contacts to bulky leads~(Fig.~\ref{fig:system}) made of normal
metal. The effects we study can be observed in various realizations
but we consider the simplest three-terminal version when one of the
leads is grounded, and the voltage $V$ is symmetrically applied to the two
other leads. We examine a 2D TI with the helical edge
state~(Fig.~\ref{fig:system-edge}) and a 3D TI cylinder with a
conducting surface state~(Fig.~\ref{fig:system-surf}). We denote the
length of the tunnel contact to the grounded lead by $l_0$, while
$l_1$ and $l_2$ stand for the lengths of the contacts to the leads
with potentials $V_{\pm} = \pm V/2$.

The total Hamiltonian reads
\begin{equation}
  \hat H = \hat H_{TI} + \sum\limits_{i=1,2,3} \hat H_{lead,i}+ \hat H_{tun,i}.
  \label{eqn:total-Hamiltonian}
\end{equation}
Here $\hat H_{lead,i}$ is the Hamiltonian of the $i$-th lead, $\hat
H_{TI}$ is the Hamiltonian of the conducting state in TI. For the edge
state\cite{KonigJPhysSocJpn2008,ZhangNaturePhys2009,HasanKane2010colloqium}
\begin{equation}
  \hat H^{(edge)}_{TI} = \int dx \hat \Psi^\dag(x)\left(-i\sigma_z v_F \partial_x - \varepsilon_F \right)\hat \Psi(x),
  \label{eqn:edge-Hamiltonian}
\end{equation}
where $v_F$ is the velocity of the excitations, $\hat \Psi$ is a
two-component spinor and $\boldsymbol{\sigma}$ are the Pauli
matrices. We do not take into account impurity scattering in the 2D
case, since spin-momentum locking prohibits such a scattering.  For
the surface state the Hamiltonian reads in the simplest
case~\cite{HasanKane2010colloqium, ZhangNaturePhys2009}
\begin{equation}
  \hat H^{(surf)}_{TI} = \int d^2\mathbf{r}\; \hat \Psi^\dag(\mathbf{r})\left[ \left( -iv_F\partial_{\mathbf{r}}\times \mathbf{e}_z \cdot \boldsymbol{\sigma}  \right) -\varepsilon_F  + V_{imp}(\mathbf{r})\right]\hat\Psi(\mathbf{r}),
  \label{eqn:surf-Hamiltonian}
\end{equation}
where $\mathbf{e}_z$ is a unit vector perpendicular to the surface,
$V_{imp}$ is a random potential of impurities, and we assume that it
is delta-correlated $\overline{V(\mathbf{r}) V(\mathbf{r}')}=u_0
\delta(\mathbf{r}-\mathbf{r}') $.

The tunnel Hamiltonian $\hat H_{tun}$ reads
\begin{equation}
  \hat H_{tun} = \int d^3 \mathbf{R} d^D \mathbf{r}\; \hat \psi^\dag(\mathbf{R}) \mathcal{T}(R,r) \hat \Psi(\mathbf{r}) + h.c.
  \label{eqn:tunnel-Hamiltonian}
\end{equation}
where dimension $D=1$ for the edge state and $D=2$ for the surface
state; $\hat \psi(\mathbf{R})$ is the field operator in a lead, the
matrix element $\mathcal{T}(\mathbf{R},\mathbf{r})$ describes
tunneling between the lead and TI. We assume a site-to-site tunnelling
which does not conserve momentum,
$\mathcal{T}(\mathbf{R},\mathbf{r})=td^{(3-D)/2}\delta(\mathbf{R}_{\|}-\mathbf{r})\delta(\mathbf{R}_\bot)$,
where $t$ is real and does not depend on $\mathbf{r}$, and
$\delta(\mathbf{R}_{\bot})$ selects an average value of a function at
a distance $d$ of the order of inter-atomic scale near the surface.
\begin{figure}
\end{figure}

% Tunneling into the edge state
First, we focus on the helical edge state coupled by tunnel contacts
to the leads~(Fig.~\ref{fig:system-edge}). We start from the
Hamiltonian~(\ref{eqn:total-Hamiltonian}),~(\ref{eqn:edge-Hamiltonian}),~(\ref{eqn:tunnel-Hamiltonian}),
and then derive equations for Keldysh
matrices~\cite{Keldysh1965diagram}
\begin{equation*}
  \check{G} = \begin{pmatrix}
    G^R&G^K\\
    0&G^A
  \end{pmatrix},\quad \check{\Sigma}_{tun} =
  \begin{pmatrix}
    \Sigma^R&\Sigma^K\\
    0&\Sigma^A
  \end{pmatrix},
\end{equation*}
where $G^{R,K,A}$ are Green functions of the edge state,
$\Sigma_{tun}$ is a self energy describing tunneling from a lead to the
edge state. Deriving an expression for self energy we follow Kopnin
and Melnikov\cite{KopninMelnikov2011proximity}. For details one can
also refer to~Ref.~\onlinecite{StanescuJPhys2013}, where the self
energy was derived for helical states tunnel-coupled to a
superconductor. Finally, we obtain
$\Sigma_{tun}(x,x')=\Sigma\delta(x-x')$, where

\begin{equation}
  \check{\Sigma}  = i\Gamma
  \begin{pmatrix}
    -1& -2\tanh \dfrac{\varepsilon}{2T}\\
    0&1
  \end{pmatrix}.
  \label{eqn:self-energy}
\end{equation}
Here we introduce the tunnelling rate $\Gamma \simeq \pi \nu_3 d^3 t^2
\sim t^2/\varepsilon_F$, $\nu_3 = mp_F/(2\pi^2 \hbar^3)$ is the 3D
density of states.

The Dyson equation for the Green functions $\check G$ reads
\begin{equation}
  \left(\varepsilon + \varepsilon_F + i\sigma_z v_F \partial_x - \check{\Sigma} \right)\check{G}(x,x') = \delta(x-x')
  \label{eqn:edge-Dyson}
\end{equation}
The left-right subtracted Dyson equation for $G^K(x,x)$ can be reduced
to a kinetic equation for distribution function $f$ by ansatz
$G^K=(G^R-G^A)(1-2f)$
\begin{equation}
  \sigma_z v_F \partial_x f = -\gamma(x)( f -f_i),
  \label{eqn:edge-quasiclassics}
\end{equation}
where $\gamma = 2\Gamma/ v_F$ is the inverse damping length due to
tunneling, $f_i = f_0 (\varepsilon- V_i)$ is the equilibrium
distribution function in the $i$-th lead.

Solving~(\ref{eqn:edge-Dyson}) for retarded and advanced components we
obtain
\begin{equation}
  G^R(x,x) - G^A(x,x) = -\frac{i}{v_F} \frac{\sinh \gamma l/2}{\cosh \gamma l/2 - \cos\left( k_FL  +\varepsilon L/v_F \right)}
\end{equation}
where $l=l_0+l_1+l_2$ and $L$ is the circumference of the edge
state. The solution of~(\ref{eqn:edge-quasiclassics}) can be
represented as a sum of equilibrium and non-equilibrium terms
$f=f_0+\delta
f$. %, where $f_0$ is a zero-voltage Fermi distribution function.
Non-equilibrium term at the region $0<x<l_0$ coupled to the grounded
lead reads
\begin{equation}
  \delta f = \frac{\delta f_2 + \left[\delta f_1-\delta f_2\right]e^{-\gamma\sigma_zl_2} -
    \delta f_1e^{-\gamma\sigma_z(l_1+l_2)}  }{\left(1-e^{-\gamma\sigma_z l}\right) e^{\gamma\sigma_z x}},
\end{equation}
where $\delta f_i = f_0 (\varepsilon- V_i)-f_0 (\varepsilon)$. The charge current flowing through the edge
state is related to the non-equilibrium part of the Keldysh Green
function $G^K_{ne}$ by $I_e = \frac{i}{2}e v_F\mathrm{Tr}\; \sigma_z
G^K_{ne}$. Spin current reads $J_s = v_F \rho$, where
$\rho$ is a linear density of electrons related to the Keldysh Green
function and a local shift of a chemical potential $\mu$ by equation $\rho = e\mu/(\pi v_F) -
\frac{i}{2} \mathrm{Tr}\; G^K_{ne}$. The local shift of a chemical potential $\mu$ is due to variation of electron density and obeys the Poisson
equation $\left(\partial_x^2 + \partial_{\bot}^2\right)\mu = -4\pi e
\rho \delta\left( \mathbf{r}_\bot \right) $. Finally, we obtain $
J_s(x) \sim v_F\left(-\frac{i}{2}\mathrm{Tr}\;
  G^K_{ne} \right)/(1+ \alpha) , $ $\alpha = a e^2/(\epsilon\hbar
v_F)$, where $a \sim 1$ depends on the specific geometry, and
$\epsilon$ is an ambient dielectric constant.

The spin current flowing through the grounded lead can be calculated
as the difference of the spin currents in the edge state of TI at the
endings of the contact $J_s = J_s(x=0)-J_s(x=l_0)$. Its derivative
with respect to the applied voltage at low temperatures
$T\ll \hbar v_F/L$ reads
\begin{equation}
  \frac{dJ_s}{dV} = \frac{G_0}{e} \frac{2\sinh \frac{\gamma l_0}{2}\sinh\frac{\gamma l_1}{2}\sinh\frac{\gamma l_2}{2}}{1+\alpha}\left[\frac{1}{\cosh \gamma l/2 - \cos\left( k_FL  +\frac{eVL}{2\hbar v_F} \right)} + \frac{1}{\cosh \gamma l/2 - \cos\left( k_FL  -\frac{eV L}{2\hbar v_F} \right)} \right],
\end{equation}
where $G_0=e^2/h$ is the conductance quantum. Here and below a spin current is measured in units of $\hbar/2$.  In the limit of high
temperatures $T > \hbar v_F/L$ the oscillations are washed out, and
the term in the square brackets should be substituted by $2/(\sinh
\gamma l/2)$

The electric current flowing through the grounded lead $I =
I_e(x=0)-I_e(x=l_0)$ in case of symmetrical geometry $l_1=l_2$, is
determined by conductance
$$
\frac{dI}{dV} = G_0 \frac{\sinh \frac{\gamma
    l_0}{2}\sinh \gamma l_1 \sin k_FL \sin \frac{eVL}{2\hbar
    v_F}}{\left[\cosh \gamma l/2 - \cos\left( k_FL +\frac{eVL}{2\hbar
        v_F} \right)\right]\left[\cosh \gamma l/2 - \cos\left( k_FL
      -\frac{eV L}{2\hbar v_F} \right)\right]}
$$
that oscillates with the voltage and the Fermi level position, which
can be varied by the gate voltage. At high temperatures $T > \hbar
v_F/L$ and in the limit of large damping, $\gamma l_i \gg 1$, this
term vanishes resulting in a pure spin current through the grounded
lead.

It is instructive to consider an incoherent case $\gamma l_i \gg 1$ in
more details. In this case the non-equilibrium part of the electronic
distribution at the region coupled to the grounded lead is reduced to
\begin{equation}
  \delta f = 
  \begin{pmatrix}
    \delta f_2e^{-\gamma x}&0\\
    0& \delta f_1 e^{\gamma (l_0-x)}
  \end{pmatrix}
\end{equation}
Thus, due to the spin-momentum locking, the distribution of spin-up
electrons at $x=0$ is determined by the heat bath coupled to the
region $x<0$, and the distribution of spin-down electrons at $x=l_0$ is
determined by the heat bath coupled to the region $x>l_0$. The spin
current reads $J_s = \left[1+\alpha \right]^{-1} \frac{G_0}{e}
V$ and the electric current equals zero, independent on the lengths of the
contacts.

It is interesting that the electric current between the leads
connected to a voltage source in the considered three-terminal
structure equals $I = \frac{3}{2}G_0 V$, and is different from the
current in a two-terminal setup. In the latter case we find for the
system with two tunneling contacts the same result $I = 2G_0 V$ as in
case of ballistic quantum wire attached to ideal adiabatic
contacts.

% Tunneling into the surface state

Now we consider a surface state of a 3D TI tunnel-coupled to the leads~(Fig.~\ref{fig:system-surf}). The Hamiltonian is given
by~(\ref{eqn:total-Hamiltonian}),~(\ref{eqn:surf-Hamiltonian}),~(\ref{eqn:tunnel-Hamiltonian}). We
assume that the contacts are placed on the (111) plane -- in this case
Pauli matrices in the Hamiltonian~(\ref{eqn:surf-Hamiltonian})
coincide with the electron's spin
operator~\cite{SilvestrovPRB2012}. The Dyson equation reads
\begin{equation}
  \left[i\partial_t + \varepsilon_F + i v_F \left(\partial_y \sigma_x  -\partial_x\sigma_y \right) - \check{\Sigma}_{tun}-\check{\Sigma}_{imp} \right]\check{G}(\mathbf{r},\mathbf{r'}) = \delta(\mathbf{r}-\mathbf{r'}).
  \label{eqn:surf-Dyson}
\end{equation}
The self-energy for impurities $\check{\Sigma}_{imp} = -i\tau^{-1}
\langle\check{g}\rangle$, where $\tau^{-1} = \pi \nu_2 u^2$,
$\nu_2=p_F/(2\pi\hbar^2 v_F)$ is the single-particle density of states at the
Fermi energy and $\langle \check{g} \rangle$ is the average over the momentum direction of the
quasiclassical Green function given by definition $\check{g} =
\frac{i}{\pi}\int \check{G} d\xi$. Similarly to the case of tunneling into the edge state, the self-energy $\check\Sigma(\mathbf{r},\mathbf{r}') =
\check\Sigma(\mathbf{r})\delta(\mathbf{r}-\mathbf{r'})$, and
$\check\Sigma(\mathbf{r})$ is given by~(\ref{eqn:self-energy}). The electron
transport is determined by an interplay between three characteristic
scales: the lengths of tunnel contact $l_i$, the damping length due to
tunnel contacts $v_F/\Gamma$, and the momentum and spin relaxation
length $v_F\tau$ due to impurity scattering. We focus on the case when
the dimensions of the sample are larger than the mean free path and
the dimensional quantization can be ignored. We follow~Ref.~\onlinecite{SchwabEPL2011} and obtain the
kinetic equation for the distribution function $f$
\begin{equation}
  \partial_t f + v_F\left(\mathbf{n} , \nabla \right)f = -\frac{f-\left\langle f \right\rangle -\left(\mathbf{n},\left\langle f\mathbf{n}  \right\rangle\right)}{\tau}-
  2\Gamma(x) (f-f_i)
  \label{eqn:3D-kinetic}
\end{equation}
The distribution function $f$ yields the quasiclassical Keldysh Green function by relation
$$
g^K = (g^R-g^A)(1-2f)=\left(1 + n_y \sigma_x - n_x \sigma_y
\right)(1-2f),
$$
We represent the distribution function as a sum
of isotropic and anisotropic terms, expanding angular dependence to the first harmonics $f\approx \langle f \rangle+f_xn_x+f_yn_y$, and
the term with $n_y$ vanishes due to translational symmetry along the
$y$-axis.  The electron and current density read
\begin{equation}
  \rho=\frac{\nu_2}{2} \int \langle f \rangle
  d\varepsilon + \nu_2 \mu,\;
  j=\frac{v_F
    \nu_2}{2}\int f_x d\varepsilon,
  \label{eqn:3D-j-rho}
\end{equation}
%Then we average kinetic equation~(\ref{eqn:3D-kinetic}) and
%kinetic equation~(\ref{eqn:3D-kinetic}) multipled by $n_x$ over the
%angles and integrate over energy. In this way using
%expressions~(\ref{eqn:3D-j-rho}) we obtain the continuity equation
%that takes into account tunneling and the expression for the current
Following~Ref.~\onlinecite{SchwabEPL2011} and taking into account a local shift of a chemical potential one can obtain
the continuity equation with the source describing tunneling, and the expression for the electric current 
\begin{equation}
  \partial_t \rho + \partial_x j = 2\Gamma(x)\left[ \rho - \nu_2(\mu-V_i)\right]
  \label{eqn:continuity-equation}
\end{equation}
\begin{equation}
  j = \sigma E + D\partial_x \rho
  \label{eqn:TI-current}
\end{equation}
where $V_i$ is a potential applied to the lead, $D =
v_F^2\tau/(1+4\Gamma\tau)$, $\sigma = e^2 v_F p_F
/[2\pi(\tau^{-1}+4\Gamma-2i\omega)]$.  The spin current density in the
TI reads
\begin{equation}
  j_s =
  v_F \frac{\rho}{2}
  \label{eqn:spin-current-density}
\end{equation}
However, spin relaxation due to scattering on impurities results in
non-conservation of the spin current, and unlike the case of the edge
state we cannot calculate the spin current flowing through the lead as
the difference of the spin currents in the surface state of TI at the
endings of the contact unless the contact length $l_0$ is shorter than
the mean free path $\tau v_F$. Thus, in order to calculate the spin
current through the lead we use the continuity equation in the lead\cite{SchwabEPL2011}
\begin{equation}
  \partial_t \rho_s^{(lead)}(x,y) + \mathrm{div} j_s^{(lead)}(x,y)  = \Gamma'(x) \delta(z) \rho_s^{(lead)}(x,y)  + 2v_F^{-1}\Gamma(x) j_e^{(TI)}(x,y)\delta(z),
  \label{eqn:3D-cont}
\end{equation}
where $\Gamma' = \Gamma \nu_2 / \nu_3$, $\rho_s$ and $j_s$ are spin and spin current densities in the
lead, $j_e^{(TI)}$ is particle current density in the TI. The term
with $\rho_s$ in the right-hand side vanishes in the leading
approximation.  Integrating~(\ref{eqn:3D-cont}) over space allows us
to relate the spin current in the lead with the electric current in
the TI
\begin{equation}
  J_s =  \frac{1}{v_F}\int 2\Gamma j_e^{(TI)}dx dy
  \label{eqn:spin-electric}
\end{equation}
Note that in the limiting case $\tau^{-1}\ll \Gamma$ according
to~(\ref{eqn:TI-current})--(\ref{eqn:spin-current-density})
expression~(\ref{eqn:spin-electric}) is reduced to the difference of
the spin currents in the surface state of TI at the endings of the
contact.

Now it is straightforward to calculate the spin current through the grounded lead
using equations~(\ref{eqn:continuity-equation})--(\ref{eqn:TI-current})
and demanding continuity of particle and current densities at the
boundaries of the contacts. The result has especially simple form when
$v_F / \Gamma \ll l_i$:
$$
J_s = \frac{G_0}{e}k_F
L_y\frac{1}{\left[4+(\Gamma\tau)^{-1}\right] \left(1+sl_D \right) }eV.
$$
where $s$ is a spacing between contacts, $l_D=\sqrt{D/(8\Gamma)}$ is a
diffusion length.

The electric current through the grounded lead equals zero. If the
mean free path $\tau v_F$ is greater than the damping length due to
tunneling $v_F/\Gamma$ then impurity scattering and corresponding spin
relaxation do not affect spin current.

% Conclusion

To summarize, we have proposed a system based on the 2D/3D TI which
injects pure spin current into an external circuit. We have found that
charge and spin transport is strongly affected by contacts
connecting the TI to bulky leads which play a role of a heat bath. If
the tunneling rate is large enough so that the exchange of electrons
between the TI and the lead is intensive enough, the distribution
functions of the electrons that passed the contact are determined by
the Fermi distribution in the lead shifted by the applied
voltage. This is somewhat similar to the case of quantum wire
connected to the leads by ideal contacts, and similarly to the quantum
wires yields electric and spin currents through a 1D channel being
proportional to the conductance quantum. In case of 2D conducting
region the current through the width of the order of the Fermi
wavelength is proportional to the conductance quantum. Thus the
conductance does not depend on the transmission of the contact if the
tunnel coupling is not too weak, and the contact behaves as if it is nearly
ideal. Though formally our results are valid in the tunneling limit
only, we believe that they provide a qualitative description of the
transport for any contacts.

The work was supported by Programs of Russian Academy of Sciences and
by Fund of non-profit programs.  ''Dynasty``.

\end{document}